# The Hybrid Service Model of Electronic Resources Access in the Cloud-Based Learning Environment

Mariya Shyshkina

Institute of Information Technologies and Learning Tools of the National Academy of Pedagogical Sciences of Ukraine, Berlinskii Str., 9, Kyiv, Ukraine
marple@ukr.net

**Abstract.** Nowadays, the search for innovative technological solutions to the organization of access to electronic learning resources in the university and their configuration within the environment to fit the needs of users and to improve learning outcomes has become key issues. These solutions are based on the emerging tools among which cloud computing and ICT outsourcing have become very promising and important trends in research. The problems of providing access to electronic learning resources on the basis of cloud computing are the focus of the article. The article outlines the conceptual framework of the study by reviewing existing approaches and models for the cloud-based learning environment's architecture and design, including its advantages and disadvantages, and the features of its pedagogical application and the experience of it. The hybrid service model of access to learning resources within the university environment is described and proved. An empirical estimation of the proposed approach and current developments in its implementation are provided.

**Keywords:** hybrid model, learning environment, cloud computing, university.

**Key Terms:** ICTInfrastructure, Model, TeachingProcess

## 1 Introduction

Progress in the area of ICT and network technology has provided new insights into the problems of the formation and development of the educational environment of the university, showing a need for advanced ICT access, especially with regard to the use of the cloud-based tools and resources. There is a need for modernization of learning technologies, supported by emerging ICT, on the basis of advanced network infrastructures.

Cloud computing technology (CC) enhances multiple access and joint use of educational resources at different levels and domains. On the basis of this technology, it is possible to combine the corporate resources of the university within a united framework. To achieve this aim, a set of cloud-based learning models should be created for the elaboration and design of learning resources and the learning environment architecture to deliver access to learning resources.

*The purpose of the article* is analyse the current trends in the university cloud-based learning environment formation from the perspective of different service models used, and to substantiate and validate the hybrid service model of access to the learning resources.

The *research method* involved analysing the current research (including the domestic and foreign experience of the application of cloud-based learning services to reveal the concept of the investigation), examining existing models and approaches, estimating the current state of research development, considering existing technological solutions and psychological and pedagogical assumptions about better ways of introducing innovative technology, and conducting pedagogical experiments, surveys and expert evaluations.

## 2  Problem Statement

The progress of ICT has a significant impact on the formation of the educational environment of the university bringing with it new models of the organization of learning activity which arise on the basis of decisions about innovative technology. In this regard, the phenomenon of the cloud-based learning environment has come to the forefront as it has many progressive features including better adaptability and mobility, as well as full-scale interactivity, free network access, a unified infrastructure among others [4, 19, 20].

The challenges of making the information technology infrastructure of the university setting fit the needs of its users, taking maximum advantage of modern network technologies, and ensuring the best pedagogical outcomes to increase the learning results, has led to the search for the most reasonable ways of organizing tools and services within the framework of this environment. For this purpose, the modelling and analysis of its structure and functions, and determining the possible types and forms of learning activity in the organization have come to the fore. Among the priority issues for ICT infrastructure design is the access to software and electronic educational resources provision [4]. To choose the best solution there is a need to consider existing approaches and models to reveal possible ways of service deployment, and to analyse the existing experience of its use.

## 3  State of the Art

According to the recent research [4, 9, 13, 18, 19], the problems of implementing cloud technologies in educational institutions so as to provide software access, support collaborative learning, implement scientific and educational activities, support research and project development, exchange experience and are especially challenging. The formation of the cloud-based learning environment is recognized as a priority by the international educational community [16], and is now being intensively developed in different areas of education, including mathematics and engineering [2, 8, 11, 25, 27].

The transformation of the modern educational environment of the university by the use of the cloud-based services and cloud computing (CC) delivery platforms is an important trend in research. The topics of software virtualization and the forming of a unified ICT infrastructure on the basis of CC have become increasingly popular lines of research [8, 18, 23]. The problems with the use of private and public cloud services, their advantages and disadvantages, perspectives on their application, and targets and implementation strategies are within the spectrum of this research [7, 8, 25].

There is a gradual shift towards the outsourcing of ICT services that is likely to provide more flexible, powerful and high-quality educational services and resources [4]. There is a tendency towards the increasing use of the software-as-a-service (SaaS) tool. Along with SaaS the network design and operation, security operations, desktop computing support, datacentre provision and other services are increasingly being outsourced as well. Indeed, the use of the outsourcing mechanism for a non-core activity of any organization, as the recent surveys have observed happening in business, is now being extended into the education sector [9]. So, the study of the best practices in the use of cloud services in an educational environment, the analysis and evaluation of possible ways of development, and service quality estimation in this context have to be considered.

The valuable experience of the Massachusetts institute of technology (MIT) should be noted in concern to the cloud based learning environment formation in particular as for access to mathematical software. The Math software is available in the corporate cloud of the University for the most popular packages such as *Mathematica, Mathlab, Maple, R, Maxima* [27]. This software is delivered in the distributed mode on-line through the corporate access point. This is to save on license pay and also on computing facilities. The mathematics applications require powerful processing so it is advisable to use it in the cloud. On the other case the market need in such tools inspires its supply by the SaaS model. This is evidenced by the emergence of the cloud versions for such products as Sage MathCloud, Maple Net, MATLAB web-server, WebMathematica, Calculation Laboratory and others [2, 8]. Really there is a shift toward the cloud-based models as from the side of educational and scientific community, and also from the side of product suppliers. The learning software actually becomes a service in any case, let it be a public or a corporate cloud.

There are many disciplines where it is necessary to outsource the processing capacity: for example, the computer design for handling vast amounts of data for graphics or video applications. This is also a useful tool used to support the collaborative work of developers, as the modern graphical applications appear to be super-powerful and require joint efforts [7]. There is a research trend connected to the virtual computing laboratories (VCL) [14, 26] delivered in the cloud-based paradigm. This trend is inherent in the field of informatics, and learning resources for processing and sharing are needed.

Nowadays there are various universal cloud consumer applications, in particular MicrosoftOffice 365, Google Docs and others which gain an appropriate use in educational process [9, 23]. There is also a wide range of cloud services such as online photo and video editors, web pages processors, services for translation, check spelling, anti plagiarism and many others which are now available [23].

There is a principal transformation of approaches in concern to services supply within the cloud based infrastructure. It is considered to be a new stage of the service oriented models development [10, 24]. There is a branch of research devoted to the service oriented infrastructure in this actual perspective. The issues of service oriented architecture development and are described in [10]. The problem of turning software into a service is also posed [24]. For example, more powerful approaches for services integration appear while services compositions are used as building blocks in a process of elaboration of programming code [14]. The CC development brought the term the *service orchestration* into scientific discussion while number of web services can be combined to perform the higher level business process to manage and coordinate execution of the component processes [12]. In this regard the notion of the global software development (GSD) is considered as novel trends overcome geographical limits [12]. There is a significant revise of approaches to ICT services elaboration and this is concerned to its integration and composition.

An essential feature of the cloud computing conception is dynamical supply of computing resources, software and hardware its flexible configuration according to user needs. Due to this approach, access to educational software set on a cloud server or in a public cloud is organized. So comparison of different approaches and cloud models of software access is the current subject matter of educational research [7, 8, 23, 25]. Also the problems of quality criteria for software choice in the learning complexes to be implemented in a cloud arise. Despite of the fact that the sphere of CC is rather emerging there is a need of some comparison of the achieved experience to consider future prospects [23].

Another set of problems is concerned with the hybrid service models and infrastructure solutions combining different public and corporate services on the united platform. This trend is now especially promising for the sphere of education [8, 17]. The challenge regarding novel technological solutions and their impact guide the search for the most reasonable method of implementation.

Thus, in view of the current tendencies, the research questions are: how can we take maximum advantage of modern network technologies and compose the tools and services of the learning environment to achieve better results? What are the best ways to access electronic resources if the environment is designed mainly and essentially on the basis of CC? This brings the problem of cloud-based services modelling, integration and design to the forefront.

## 4  Pedagogical Aspects of Electronic Resources Delivery and Indicators of Research

Cloud computing technology is now one of the leading trends in the formation of the information society. It constitutes an innovative learning concept and its implementation significantly affects the content and form of different types of activities in the sphere of education [4, 13, 18].

Along with the emergence of cloud computing, the number of objects, developments and domain applications are continually growing, which indicates the rapid spread of the innovation [20]. The concept of *the cloud-based learning environment* is now in line with the wider trend; that is to say, the ICT environment of the university, where some didactic functions as well as some fundamentally important functions of scientific research are supported by the appropriately coordinated and integrated use of cloud services [20]. The *aim* of the cloud-based learning environment formation is to meet the users' educational needs. To do this, the introduction of cloud technology in the learning process should to be holistic and carried out according to the principles of *open education*, including meeting the following needs: the mobility of students and teachers, equal access to educational systems, providing qualitative education, and forming and structuring of educational services [3, 20].

The main elements of the cloud computing conception, including the types, application service models, essential features, ICT architecture and others, are reflected in the structure of the modern educational organizational systems [5]. Therefore, a number of concepts and principles that characterize the development and application of CC-based services are significant in the consideration of the educational environment design.

The concept of *electronic educational (learning) resources* (EER) appears to be the centre of attention. In particular, at the Institute of Information Technologies and Learning Tools of the National Academy of Pedagogical Sciences of Ukraine the conception that provided the definition of electronic educational resources (EER) its classification, and the ways it can be applied has been developed and proposed [5].

According to the definition given in [5, p.3], "The electronic educational resources are a kind of educational tool (for training, etc.) that are electronically placed and served in educational system data storage devices which are a set of electronic information objects (documents, documented information and instructions, information materials, procedural models, etc.)".

The elaboration of the electronic learning resources should be considered as a specific activity, which is linked to the mandatory need to take into consideration the psychological and pedagogical aspects of building an educational system methodology, the design of an open computer-based learning environment, and the involvement of the scientific and pedagogical staff, including the best teachers and educators [4].

*Cloud Service* – is a service that makes software applications, data storage or computing capacity available to users over the Internet [16]. These services are used to supply the electronic educational resources that make up the substance of a cloud-based environment, and to provide the processes of elaboration and use of the educational services.

Electronic resources appear to be both the objects and the tools of activity for a learner; therefore, these resources are used to maintain certain functions that are realized in the learning process. By the *educational service* we mean a service provided at the request (in response to an inquiry etc.) of a user that meets some service function carried out by the organization or institution (service provider, outsourcer) [4].

Nowadays, the various types of electronic educational resources that may be delivered by the cloud in the learning environment of the university constitute libraries and depositories of EER or those retrieved when open analytical information systems are used. The EER supports different types of learning and research activities, such as theoretical material studies, the search for useful information, translation and grammar checking, task solutions, testing, training, simulation, making experiments and others.

Along with the development of information and communication technologies for education, the ways and tools of access to electronic resources have changed in an evolutionarily way and its custom properties have improved. There are new types of EER supplied by means of cloud technologies. The EER of the public cloud can occupy the role of software for general purposes such as office applications, systems support processes for communication and data exchange and others, and also the special software designed for educational use [13, 23]. The number of EERs is increasing and this trend is likely to intensify. By means of CC-based tools, a significant lifting of restrictions on the implementation of access to qualitative leaning resources may be achieved. Now, these questions are not a matter of future perspective, they need practical implementation. For this purpose, the problem of the design and delivery of electronic educational resources in the cloud-based environment is a complex one and not only should technological needs be considered, but also the pedagogical aspects.

With the advance of ICT, CC technologies appear to be a factor in the change in the content, methods and organizational forms of learning and development of the open education models. Now, cloud computing technology is used to improve the educational process through the presentation of a modern learning content adequate to the goals set, the quality monitoring and evaluation of learning results at the various stages, the creation of new organizational forms of learning, the creation of innovative educational and scientific resources and electronic systems and their implementation in the process of students' self-study and classroom study, advances in computer-aided and mixed models of training and so on [20].

As noted in [5], the necessary measures for the development of the human resources' component of the software industry created in Ukraine that concern the organization of EER access in the educational institutions are as follows: improvement to EER quality, scientific-methodological research on the implementation of innovative technologies and prospective models and methods in education, the development of the normative regulatory framework, strengthening the firms and companies in the IT industry and their participation in providing educational hardware and software and so on.

Due to the significant educational potential and novel approaches to environmental design, its formation and development, these questions remain the matter of theoretical and experimental studies, the refinement of approaches, and the search for models, methods and techniques, as well as possible ways of implementation [4].

To carry out research and experimental activities and the implementation and dissemination of the results, the Joint research laboratory of the Institute of Information Technologies and Learning Tools of the NAPS of Ukraine and the Kherson State University was created in 2011 with the focus on issues of educational quality management using ICT [29].

As part of the programme of joint research work, the Kherson State University was approved as an experimental base for research on the definition and experimental verification of the didactic requirements and methods of evaluating the quality of electronic learning resources in the educational processes of the pilot schools [29]. The purpose of the experiment carried out was to identify and experimentally verify the requirements and methods of evaluating the quality of the electronic learning resources used in the educational process in secondary schools [29].

The quality evaluation of EER in the cloud-based learning environment is a separate line of work in the Laboratory's research. In this case, there are different approaches and indicators. The access organization has been changed so the models of learning activity have been changed also. There are the following questions: What features and properties have to be checked so as to measure the pedagogical effect of the cloud-based approach? With regard to the pedagogical innovation, what are the factors influencing pedagogical systems, their structure and organization? Is the improvement in learning results achieved due to the cloud-based models? In this context, the quality of EER is a criterion for estimating the level of organization and functioning of the cloud-based learning environment.

With regard to this, the following *hypothesis* is to be posed: the design of the learning environment on the basis of cloud models of access to learning resources contributes to the improvement of the quality of these resources and the improvement of the processes in this environment and their organization and functioning, resulting in an improvement in learning results.

In the cloud-based learning environment, new ways of EER quality control arise. There are specific forms of the organization of learning activity related to quality estimation. For example there are e-learning systems based on the modelling and tracking of individual trajectories of each student's progress, knowledge level and further development [28]. This presupposes the adjustment, coordination of training, consideration of pace of training, diagnosis of achieved level of mastery of the material, consideration of a broad range of various facilities for learning to ensure suitability for a larger contingent of users. The vast data collections about the students' rates of learning are situated and processed in the cloud [28]. There are also collaborative forms of learning where the students and teachers take part in the process of resource elaboration and assessment; this is possible in particular by means of the SageMathCloud platform [2].

The prospective way of the estimation of the quality of learning resources is by means of the cloud-based environment. As the resources are collectively accessed, there is a way to allow experts into the learning process so they may observe and research their functioning. This is a way to make the process of quality estimation easier, more flexible and quicker. The process of estimation becomes anticipatory and timely. The estimation may be obtained just once along with the process of EER elaboration, and it is very important to facilitate the process.

This method of estimation was developed and used in the Joint laboratory of EER quality control [29]. In this case, the different quality parameters will be detailed and selected. It is important that the psychological and pedagogical parameters are estimated in the experimental learning process, while the other types of parameter such as technological or ergonomic may be estimated out of this process.

The indicator of accessibility is also included in the focus of this investigation [15]. This property is essential because it is prior to other features such as scientific correctness, clarity, consistency and others, which may be researched only if this resource is available and feasible. The accessibility is characterized in turn by such features as convenience of the access organization, ease of use, interface consistency, advisability and others.

## 5 The Types of Service Models for Learning Resources Access

According to recent research, a *unified storage architecture* is an advantage of cloud based settings allowing application virtualization [18, 19]. This architecture is designed for the large complex data sets retrieving and management and it has the following features:
- different storage protocols are maintained in the same system (FC, NFS, FcoE, CIFS, iSCSI);
- various storage functions are implemented within the same device (storage, security, backup, recovery);
- storage space is scaled and modified without interruption of usual operations;
- data are integrated in a standard pool, which can be controlled over a network and managed via standard software package;
- data are used for different range of applications while storage area is not necessarily separated to enable saving computing capacity through virtualization.

*Application virtualization* is a technology for software access and development without installing it on a personal computer of a user. Data processing and storage is fulfilled in a data centre, and working with applications is not different for a customer from the working with applications installed on his (her) own desktop.

There are three main types of *service models* [16] that correspond to different ways the ICT outsourcing used to provide software and computing resources access [4]. In particular, *SaaS* (Software-as a Service) is to deliver software applications of a provider via the Internet; *PaaS* (Platform as a Service) is to develop and implement software applications created by a user via the Internet; *IaaS* (Infrastructure as a Service) is to provide on-line infrastructure where a customer may develop whatever software applications [19].

P.Mell and T.Grance define various service models of the cloud-based architecture (Fig.1) [16]. These models may be purposefully used for providing software access in educational institutions.

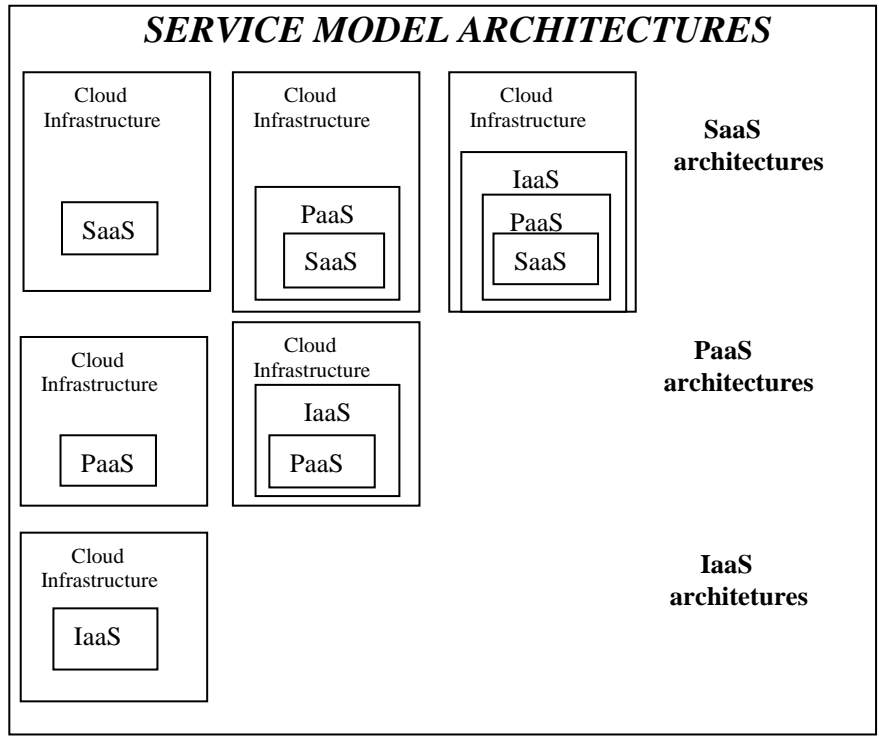

**Fig. 1.** Service model architectures (After P.Mell, T.Grance [16]).

There are also four *service deployment models* for cloud computing application that reflect the mode of the cloud infrastructure set up in a particular organization: *the corporate cloud* is owned or leased by the organization; *the cloud community* is a shared infrastructure used by a community; *the public cloud* is a mega-scale infrastructure that may be used by any person under some payment terms; *the hybrid cloud* is a composition of one or more models [4, 16].

In view of the different models for cloud service architecture, when choosing the most appropriate solution that is suitable for a particular organisation, both collective and individual users should be considered. Selecting the SaaS model in this respect can be justified by the fact that these services are the most accessible, although a thorough market analysis and educationally prudent choice of the necessary application that could fit learning or scientific purposes is to be made. This kind of service may be purposefully used by an individual and also a collective user.

At the same time, for the settlement of the ICT infrastructure of the institution by the PaaS or IaaS model, the selection and approval of the relevant cloud platform is necessary. This solution is concerned with a number of organisational issues, such as the formation of a special unit of ICT personnel skilled in setting up and deploying this infrastructure, configuring the hardware and software complexes, planning and working out the environmental design tasks, testing and approving its modules and components, filling it with the necessary resources, monitoring its implementation, maintaining quality control, training the teaching staff, etc. [4]. In this case, given the results of recent research and the current trends in IT sector development, the use of hybrid service models appears to be a promising and prospective solution. The hybrid solutions are reported to be well-embedded into existing settlements provided by leading cloud suppliers, and this tendency is growing [9]. The hybrid cloud incorporates public and corporate cloud tools that do not necessarily exclude the involvement of software-as-a-service applications [19].

As shown in Fig.1, there are three approaches to implementation of learning software access in the SaaS architecture. In the first case (directly SaaS) the cloud platform deployment is not necessary in the educational institution this work is undertaken by a service provider. In both other cases the corporate or hybrid cloud deployment is needed. In this case the appropriate cloud platform (eg, Amazon Web Services, Microsoft Azure, Eucaliptus, Xen, WMWare etc.) is used to deploy the certain service model. In the process of cloud infrastructure configuration the guidelines are usually supplied by the vendor [1]. These guidelines contain a number of basic deployment scenarios that can be implemented. It is possible to build the cloud by means of different software and services but the basic notions are to be considered. One of the basic concepts of the cloud based learning environment configuration is the concept of a *corporate cloud* or a *virtual private cloud* (VPC). Sometimes the term is used not very clearly so as to describe the corporate cloud that may include a public and also a private part so being the hybrid one. Depending on the scenario chosen the certain model of software access is considered.

As a rule the cloud provider may propose services of several types. For example, it is possible to rent additional disk space (S3); the virtual machine (EC2), with certain parameters of the processor, memory, and disk capacity, it may be with some installed operational system and software; remote database (SimpleBD, RDS) and others [1]. Depending on the chosen scenario these resources are configured within the cloud infrastructure.

There are four types of scenario for the cloud infrastructure configuration that are mostly proposed by the provider [1]:

*Scenario 1:* VPC with only public subnet. The configuration of the virtual cloud under this scenario contains a single public subnet and Internet gateway so as to enable communication over the Internet. This configuration is recommended if it is necessary to run the single level, public web applications such as blogs, web sites [1].

*Scenario 2:* VPC with public and corporate subnets. The configuration for this scenario includes a public and corporate (private) subnet. This configuration is recommended if it is necessary to run a public web application, while internal servers are not publicly available. An example is a multi-website with the web servers situated in a public subnet, and the database servers to be in a corporate subnet. It is possible to configure the security services and routing so that the web servers could interact with the database servers [1].

*Scenario 3:* VPC with the public and corporate subnet components and virtual private network (VPN) access. The configuration for this scenario includes the virtual hybrid cloud with the public and corporate subnets and the virtual corporate gateway namely the VPN connections. In an educational institution may be own subnet, which should be expanded by augmented cloud services, such as additional disk space, databases, virtual machines, network gateways, additional "desktops" and so on. VPN-connection is used to enable communication with this subnet. You can also create the virtual cloud subsystem (subnet virtual machine) with access to the corporate subnet via the Internet [1]. For this scenario, the multi-level applications with scalable web services may be run, some parts of these applications are in the public subnet, and another parts are in the corporate subnet, which is connected to own subnet through the VPN channel [1]. This allows you to keep some data in the limited access.

*Scenario 4:* VPC subnet with the corporate VPN access components. The configuration for this scenario includes the virtual corporate subnet and the virtual gateway to allow communication with own subnet through the VPN channel. This scenario is recommended if there is a need to expand own subnet into the cloud, as well as to provide direct access to the Internet from this subnet without making it "visible" from the Internet [1].

On the stage of environment design all the possible configurations were considered and the model of the Scenario 3 was chosen so as to provide the hybrid infrastructure were the corporate and public components were used (Fig. 2).

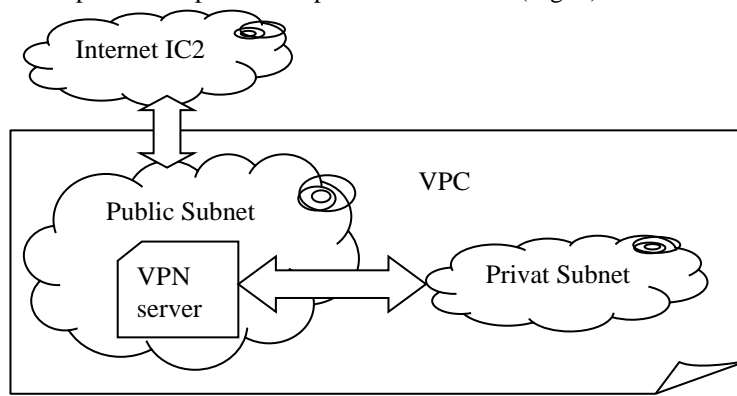

**Fig. 2.** The Hybrid Cloud configuration.

## 6   The Hybrid Service Model of Learning Software Access

To research the hybrid service model of learning software access, a joint investigation was undertaken in 2013–2014 at the Institute of Information Technologies and Learning Tools of the NAPS of Ukraine and Drohobych State Pedagogical University named after I.Franko. At the pedagogical university, the experimental base was established where the cloud version of the Maxima system (which is mathematical software), installed on a virtual server running Ubuntu 10.04 (Lucid Lynks), was implemented. In the repository of this operational system is a version of Maxima based on the editor Emacs, which was installed on a student's virtual desktop [21]. In this case, the implementation of software access due to the hybrid cloud deployment in Scenario 3 was organised.

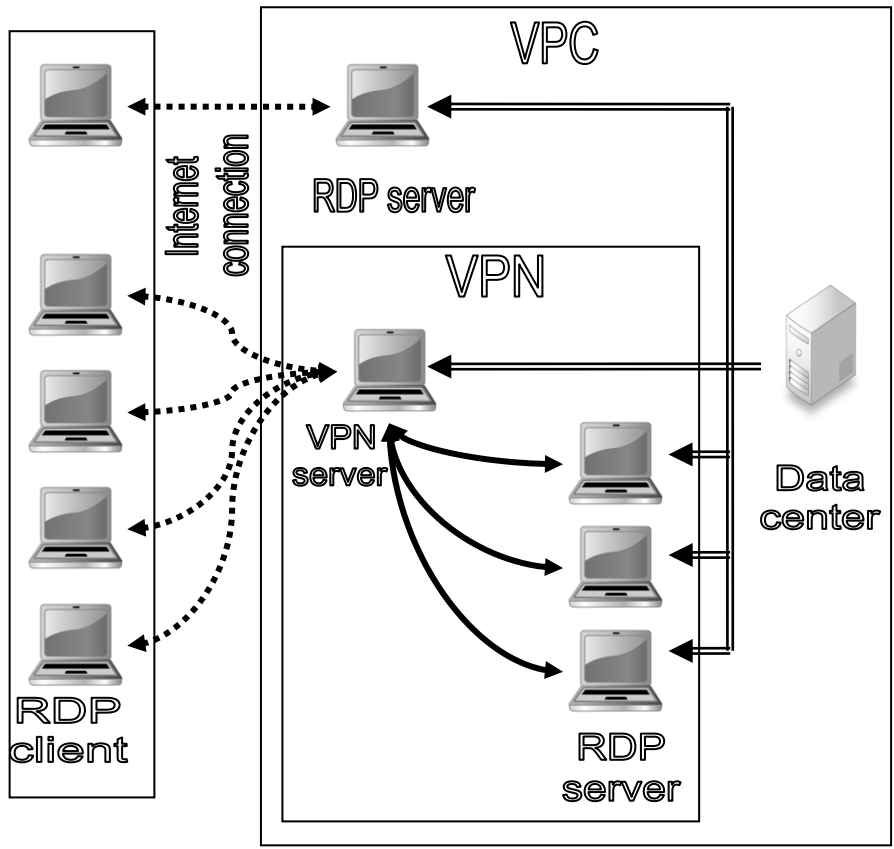

**Fig. 3.** The hybrid service model of the learning resources access.

In Fig.3, the configuration of the virtual hybrid cloud used in the pedagogical experiment is shown. The model contains a virtual corporate (private) subnet and a public subnet. The public subnet can be accessed by a user through the remote desktop protocol (RDP). In this case, a user (student) refers to certain electronic resources and a computing capacity set on a virtual machine of the cloud server from any device, anywhere and at any time, using the Internet connection.

In this case, a user's computer is the RDP-client, while the virtual machine in the cloud is the RDP-server. In the case of a corporate (private) subnet, a user cannot apply to the RDP-server via desktop because it is not connected to the Internet directly. Computers in the corporate subnet have Internet access via the VPN-connection, i.e. the gateway. Thus, these computers cannot be accessed from any device, but only from the specially configured one (for example, a computer in the educational institution or any other device where the VPN-connection is set up) (Fig.3).

The advantage of the proposed model is that, in a learning process, it is necessary to use both corporate and public learning resources for special purposes. In particular, the corporate cloud contains limited access software; this may be due to the copyright being owned by an author, or the use of licensed software products, personal data and other information of corporate use. In addition, there is a considerable saving of computational resources, as the software used in the distributed mode does not require direct Internet access for each student. At the same time, there is a possibility of placing some public resources on a virtual server so the learner can access them via the Internet and use the server with the powerful processing capabilities in any place and at any time. These resources are in the public cloud and can be supplied as needed.

## 7   Implementation and Empirical Evaluation

In the joint research experiment held at Drohobych State Pedagogical University named after I.Franko, 240 students participated. The aim was to test the specially designed learning environment for training the operations research skills on the basis of the Maxima system. During the study, the formation of students' professional competence by means of a special training method was examined. The experiment confirmed the rise of the student competence, which was shown using the $\chi^2$– Pearson criterion [21]. This result was achieved through a deepening of the research component of training. The experiment was designed using a local version of the Maxima system installed on a student's desktop.

The special aspect of the study was the expansion of these results using the cloud version of the Maxima system that was posted on a virtual desktop. In the first case study (with the local version), this tool was applied only in special training situations. In the second case study (the cloud version), the students' research activity with the system extended beyond the classroom time. This, in turn, was used to improve the learning outcomes.

The cloud-based electronic learning resource used in the experiment has undergone a quality estimation. The method of quality estimation in the joint laboratory of educational quality management with the use of ICT was used for this study [29]. The 25 experts were specially selected as having experience in teaching professional disciplines focused on the use of ICT and being involved in the evaluation process. The experts evaluated the electronic resource with such parameters as "Ease of access", "Ease of use" and "Usefulness". These parameters were chosen as they contribute to the accessibility of the cloud resource and the cloud-based learning in order to determine its feasibility and availability.

The problem was: is it reasonable and feasible to arrange the environment in a proposed way? There were three questions part of the access realisation mode (Table 1):

**Table 1.** The questionnaire.

| | Parameter | | Value |
|---|---|---|---|
| 1 | Ease of access | Is the electronic resource access easy and convenient? | 0 (no), 1 (low), 2 (good), 3 (excellent) |
| 2 | Ease of use | Is the user interface clear and convenient? | 0 (no), 1 (low), 2 (good), 3 (excellent) |
| 3 | Usefulness | Is this resource useful? | 0 (no), 1 (low), 2 (good), 3 (excellent) |

A four-point scale (0 (no), 1 (low), 2 (good), 3 (excellent)) was used for the questions. The 25 experts estimated two parameters, "Ease of access" and "Ease of use", and were invited to examine the resource. Experience using this resource in the learning process was not mandatory. The third parameter, "Usefulness", was estimated only by the seven experts who used the resource in the learning process. The results of the evaluation are shown in Fig.4.

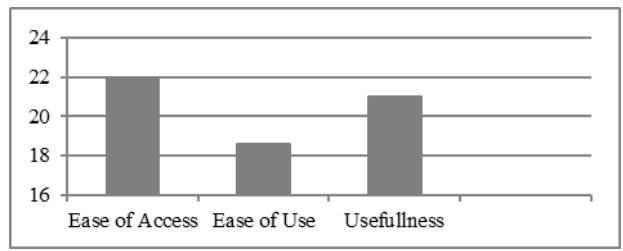

**Fig. 4.** The results of the cloud-based learning resource quality parameters evaluation.

The resulting average value was calculated for every parameter: "Ease of access" = 2.2, "Ease of use" = 1.86 and "Usefulness = 2.1. All criteria were weighted as one, and the total value was 2.1. This characterises the resource accessibility as sufficient for further implementation and use.

The advantage of the approach is the possibility to compare the different ways to implement resources with regard to the learning infrastructure. Future research in this area should consider different types of resources and environments.

## 8 Conclusion

The introduction of innovative technological solutions into the learning environment of educational institutions contributes to unified learning infrastructure formation and the growth of access to the best examples of electronic resources and services. ICT use is promising regarding learning settings that can advance and develop the tendencies of CC progress. For example, using the cloud-based models of environment design, virtualising applications, unifying infrastructure, integrating services, increasing the use of electronic resources, expanding collaborative forms of work, widening the use of the hybrid models of ICT delivery and increasing the quality of electronic resources.